\documentclass[aps,prb,preprintnumbers,showpacs,superscriptaddress,amsmath,amssymb,twocolumn]{revtex4-1}

\usepackage{silence}
\usepackage{graphicx}  
\usepackage{dcolumn}   
\usepackage{bm}        
\usepackage{amssymb}   
\usepackage{float}
\usepackage{epsfig,amsmath,amssymb,color,float,setspace}
\usepackage[utf8]{inputenc}
\usepackage[T1]{fontenc}
\usepackage{booktabs}
\usepackage{gensymb}
\usepackage[symbol]{footmisc}

\begin{document}
\singlespacing

\title{Electrical and magnetic properties of thin films of the spin-filter material CrVTiAl}

\author{Gregory M. Stephen}
\thanks{These two authors contributed equally}
\affiliation{Department of Physics, Northeastern University, Boston, MA 02115 USA}
\author{Christopher Lane}
\thanks{These two authors contributed equally}
\affiliation{Department of Physics, Northeastern University, Boston, MA 02115 USA}
\author{Gianina Buda}
\affiliation{Department of Physics, Northeastern University, Boston, MA 02115 USA}
\author{David Graf}
\affiliation{National High Magnetic Field Laboratory, Florida State University, Tallahassee, Florida 32310, USA}
\author{Stanislaw Kaprzyk}
\thanks{Deceased October 2018}
\affiliation{Faculty of Physics and Applied Computer Science, AGH University of Science and Technology, aleja Mickiewicza 30, 30-059 Krakow, Poland}
\affiliation{Department of Physics, Northeastern University, Boston, MA 02115 USA}
\author{Bernardo Barbiellini}
\affiliation{ Department of Physics, School of Engineering Science, LUT University, FI-53851 Lappeenranta, Finland}
\affiliation{Department of Physics, Northeastern University, Boston, MA 02115 USA}
\author{Arun Bansil}
\affiliation{Department of Physics, Northeastern University, Boston, MA 02115 USA}
\author{Don Heiman}
\affiliation{Department of Physics, Northeastern University, Boston, MA 02115 USA}
\affiliation{Plasma Science and Fusion Center, MIT, Cambridge, MA 02139 USA }
\date{\today}

\begin{abstract}
The spin-filter material CrVTiAl is a promising candidate for producing highly spin-polarized currents at room temperature in a nonmagnetic architecture. Thin films of compensated-ferrimagnetic CrVTiAl have been grown and their electrical and magnetic properties have been studied. The resistivity shows two-channel semiconducting behavior with one disordered gapless channel and a gapped channel with activation energy $\Delta E$=~0.1~-~0.2~eV. Magnetoresistance measurements to B~=~35~T provide values for the mobilities of the gapless channel, leading to an order of magnitude difference in the carrier effective masses, which are in reasonable accord with our density-functional-theory based results. The density of states and electronic band structure is computed for permutations of the four sublattices arranged differently along the (111) body diagonal, yielding metallic (Cr-V-Al-Ti), spin-gapless (Cr-V-Ti-Al) and spin-filtering (Cr-Ti-V-Al) phases. Robustness of the spin-gapless phase to substitutional disorder is also considered.  
\end{abstract}

\maketitle

\section{INTRODUCTION}

\begin{figure}
    \centering
    \includegraphics[width=5cm]{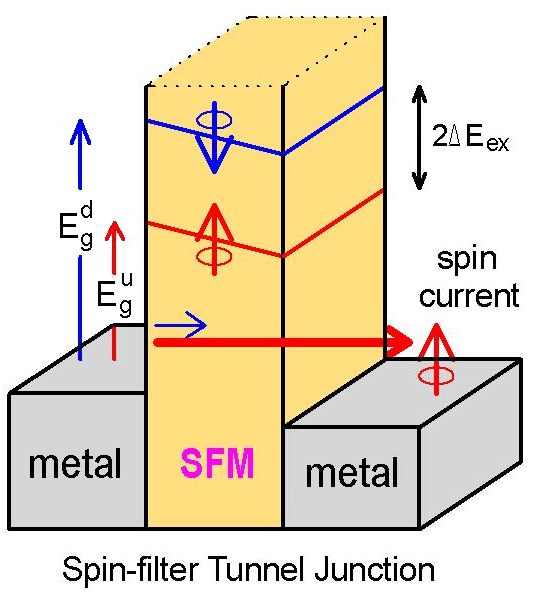}
    \caption{Schematic of a spin-filter device. The SFM is sandwiched between two nonmagnetic metallic contacts. As the barrier height for the electrons with different spin directions is different, the spin with the smaller band gap (barrier height) dominates the tunneling current.}
    \label{fig:SFM}
\end{figure}

In order to fully utilize the electron spin in spintronic devices, it is necessary to produce highly spin-polarized currents at room temperature.\cite{Wolf2001,Zutic2004} Spin-filter materials (SFMs) are semiconductors with exchange splitting $2\Delta E_{ex}$ between the two spin-channels. When employed as a tunneling barrier as shown in Fig. \ref{fig:SFM}, electrons with one spin direction can tunnel more easily through an SFM barrier to produce a spin-polarized current. Since the tunneling probability depends exponentially on the band gap, even a small exchange splitting can yield a large spin-polarization.\cite{Moodera1988} In addition, because the barrier material itself is spin-polarized, the metallic contacts need not be magnetic. Therefore, when using an antiferromagnetic or compensated ferrimagnetic SFM, the entire device can be free of fringing fields and effectively nonmagnetic.

The europium chalcogenides have been shown to produce nearly 100~\% spin-polarized currents when used as tunneling barriers.\cite{Moodera2010,Santos2004}  However, their magnetic transition temperatures are below T~=~70~K, making them ineffective for room temperature applications. Thus, it is necessary to find SFMs with high magnetic transition temperatures. 

Heusler compounds have gained increasing interest as spin-polarized materials, as they span the gauntlet of spin-dependent band structures.\cite{Felser2015,Graf2011,Bainsla2016} However, while a plethora of half-metals and spin-gapless semiconductors have been found both theoretically and experimentally among the Heusler compounds, only a handful of SFMs have been predicted (CrVXAl and CoVXAl, where X = Ti, Zr, Hf) by Galanakis et al.\cite{Galanakis2013,Galanakis2014} and Ti\textsubscript{2}CrSn by Jia. et al.\cite{Jia2014a}.) Among these Heusler SFMs, CrVTiAl is a promising candidate material that could be utilized at room temperature, as it is a fully-compensated ferrimagnet with a predicted $T_C~>~2000$ K.\cite{Galanakis2014, Ozdogan2015} Previous experiments on bulk CrVTiAl have established that $T_C$ is in excess of 400 K and that partially disordered samples show semiconducting behavior with a magnetic moment that is small and nearly temperature independent.\cite{Stephen2016, Venkateswara2017,Stephen2019a} Density functional theory (DFT) studies have revealed that the semiconducting phase of CrVTiAl is stable under Cr-V swapping, but the material becomes metallic under Ti-Al mixing. \cite{Stephen2019a}

An outline of this paper is as follows. The introductory section is followed by a discussion of the experimental results in Section II and the computational results in Section III. Section II contains three sub-sections: (A) The measured zero-field conductivity, illustrating two-channel semiconducting behavior in the films; (B) Magnetoresistance up to fields of B~=~35 T, along with fits to a two-channel conduction model and extraction of the mobilities of the two channels; and (C) Hall resistivity and its relationship to a small magnetic moment saturating by B~=~15 T. Section III contains two sub-sections: (A) Densities of states (DOSs) of the three ordered structures generated by permuting the atoms along the (111) body diagonal; and (B) effects of substitutional disorder among the sublattices. Section IV, gives a short summary of our findings

\section{EXPERIMENTAL PROCEDURES AND RESULTS}

Films of CrVTiAl were grown on Si/SiO\textsubscript{2} substrates using magnetron sputtering and capped with 1 nm of Al. Growth temperatures were in the range 300 to 600~\degree C and films were post-growth annealed in the range 500 to 700~\degree C. Films were typically 30 to 40~nm thick. This paper focuses on measurements of a film grown at 400~\degree C and subsequently annealed to 600~\degree C and 700~\degree C. The three films will be simply denoted as F400, F600 and F700, respectively.  Stoichiometry was confirmed to be within 3~\% of the ideal using energy dispersive X-ray spectroscopy. Magnetization measurements were performed in a Quantum Design MPMS-XL magnetometer at temperatures T~=~2 to 400~K and magnetic fields up to B =~5~T. Zero-field resistivity measurements were performed in the magnetometer using a modified transport probe.\cite{Assaf2012} High-field resistivity measurements up to B =~35~T were performed at the National High Magnetic Field Laboratory DC Field Facility.

\subsection{Electrical Conductivity}
\begin{figure}
\includegraphics[width=8cm]{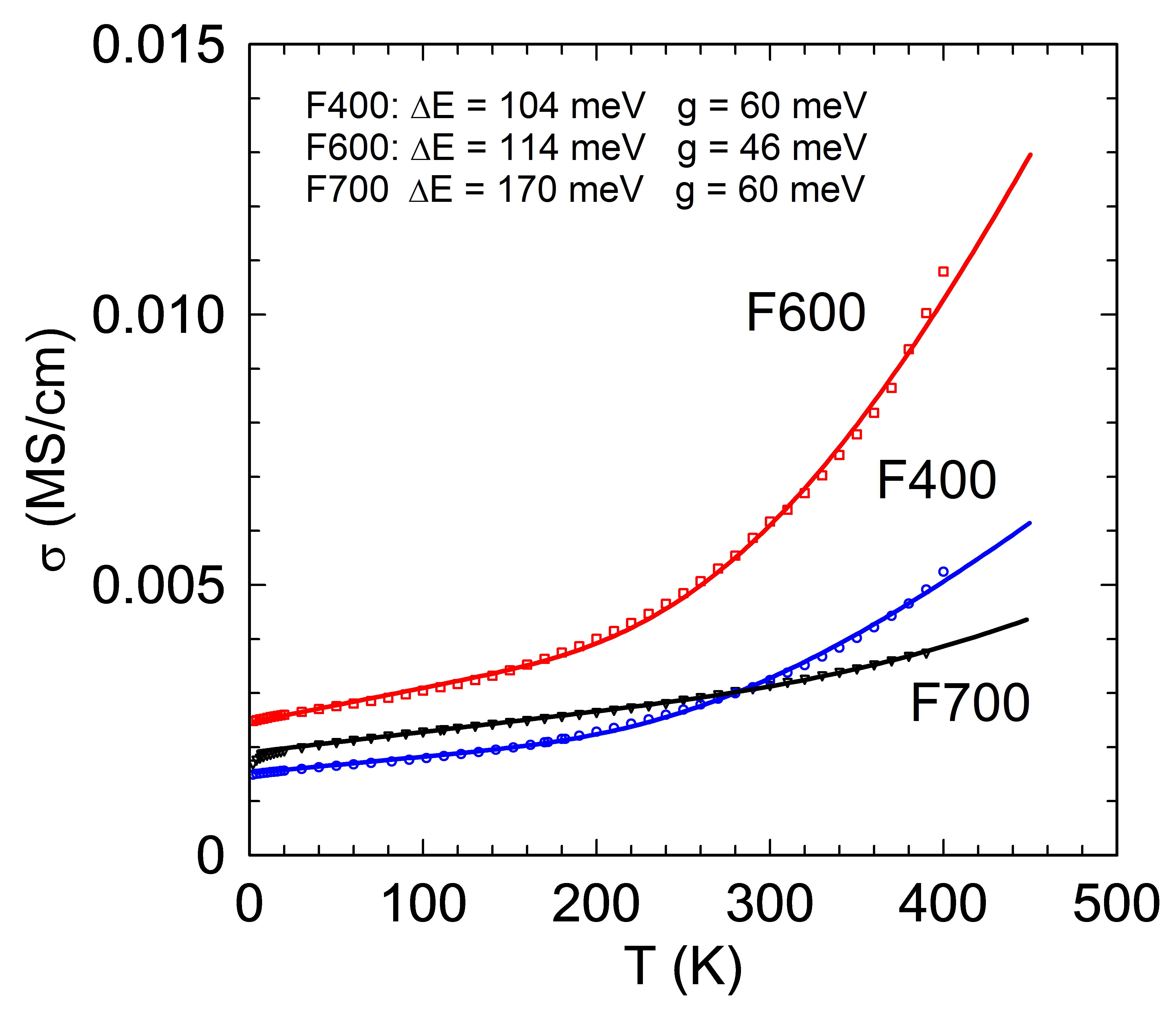}
     \caption{Zero-field conductivity versus temperature for three different films of CrVTiAl (F400, F600, and F700). The increasing conductivity indicates semiconducting behavior. The points are experimental data and the solid curves are fits to Eq. (2).}
\label{fig:RT}
\end{figure}

Figure \ref{fig:RT} shows the temperature-dependent conductivity $\sigma$(T) for the CrVTiAl thin films (points). $\sigma$(T) is observed to increase linearly for moderate temperatures characteristic of spin-gapless (SGS) behavior, while at higher temperatures it follows an exponential dependence indicative of the presence of a semiconducting gap. Modeling this temperature-dependent $\sigma$(T) provides information about carrier activation energies, which are related to semiconducting band gaps. This, in turn, allows one to infer the nature of the band structure near the Fermi energy, whether it is gapped, gapless or metallic.

The measured $\sigma$(T) dependence can be described here by a two-carrier conduction model along the lines of previous studies\cite{Wang2008,Jamer2013,Kharel2015}. The model assumes two parallel conducting channels: one gapped and one gapless. The conductivity in the gapless channel is given by
\begin{equation}
    \sigma_{SGS} = \sigma_0 (1+ 2ln(2)\frac{k_BT}{g} ),
    \label{eq:SGS_Ch}
\end{equation}
where $\sigma_{0}$ is the zero-temperature conductivity, $g$ denotes the overlap between the bands in the gapless channel and k$_{B}$ is the Boltzmann constant.\cite{Kharel2015} Combining this in parallel with a conventional semiconducting conduction channel with a thermal activation energy $\Delta E$ gives the temperature-dependent resistivity $\rho$(T)
\begin{equation}
    1/\rho(T) = \sigma(T) = \sigma_{SGS} + \sigma_{SC} e^{-\Delta E/k_BT},
\end{equation}
where $\sigma_{SC}$ is the zero-temperature contribution to the semiconducting component.

The solid curves in Fig. \ref{fig:RT} show fits to the model in Eq. (2), giving the following activation energies ($\Delta E$) and band overlaps ($g)$ for the three films. Activation energies are: $\Delta E$~=~104 meV  for F400, 114 meV for F600, and 170 meV for F700. Band overlaps are: $g$~=~60 meV for both F400 and F700, and $g$~=~46 meV for F600. The activation energy increases with increased annealing temperature, while $g$ is at a minimum for the F600 film.

\subsection{Magnetoresistance}

The magnetoresistance (MR) shows four distinct characteristics: (1) two-channel positive MR; (2) a positive weak localization cusp; (3) suppressed MR at low fields and higher temperatures; and (4) the onset of Shubnikov-de Haas quantum oscillations at higher fields. These features are discussed in the following subsections.

\begin{figure*}[t]
\includegraphics[width=2\columnwidth]{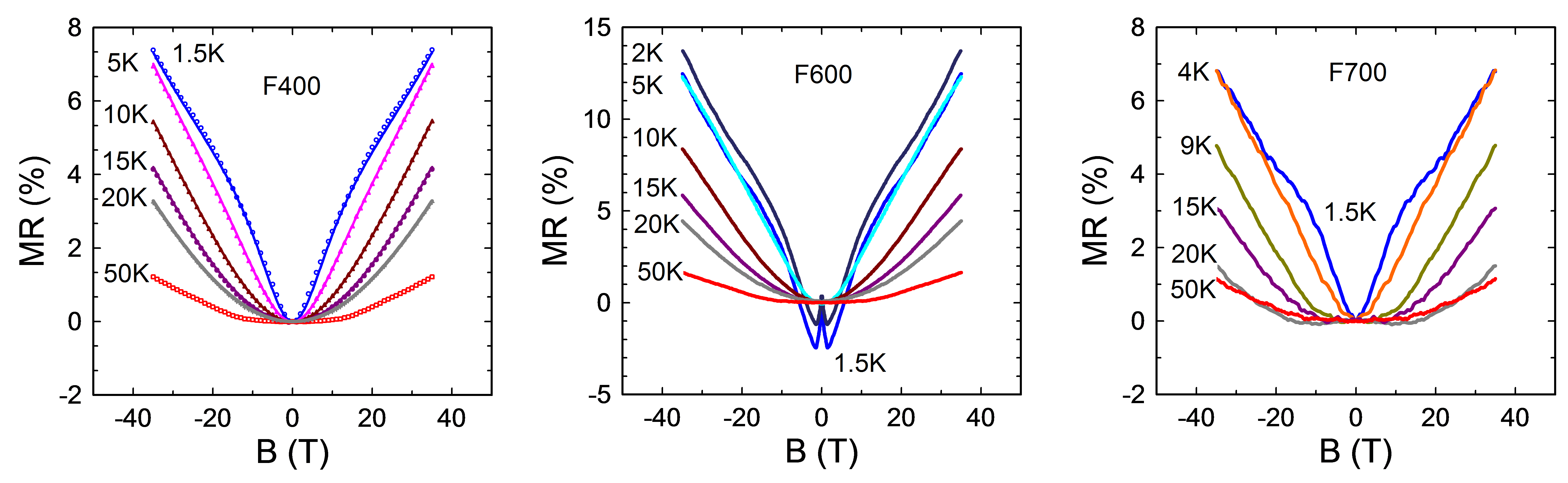}
     \caption{Magnetoresistance as a function of applied field for CrVTiAl films (labeled F400, F600, and F700; as described in the text), measured up to $B~=~35$~T at temperatures from T~=~1.5 to 50~K. The solid curves for the as-grown 400~\degree C sample (F400) for T~$\leq$~20~K are fits to Eq. (5). All three samples generally show quadratic-in-B behavior at high temperatures and linear-in-B behavior at low temperatures}
\label{fig:MR}
\end{figure*}

\subsubsection{Two-channel positive MR}

Figure \ref{fig:MR} shows the magnetoresistance as a function of magnetic field and temperature. The field dependence has an overall positive MR for different annealing conditions. The MR(B) behavior is illustrated best in the as-grown sample, Fig. \ref{fig:MR}(F400). Here, at higher temperatures the MR is predominantly quadratic-in-B, but as the temperature decreases the MR at high fields becomes linear-in-B. Then at the lowest temperatures, T~<~5~K, there is an inflection point near B~=~10~T.

The MR(B) dependence over the entire field range can be modeled as two parallel and independent conduction channels. The field-dependent conductivity associated with the $i$-th channel is given by

\begin{equation}
\frac{1}{\rho_i(B)} = \sigma_i(B) = \frac{\sigma_{i,0}}{1+\mu_i^2 B^2},
\end{equation}
where $\sigma_{i,0}$ is the zero-field conductivity and $\mu_i$ is the carrier mobility for the $i$-th channel. By combining $\rho$(B) for the two channels in parallel, and labeling light and heavy mass channels, denoted as \textit{l} and \textit{h}, respectively, the total $\rho$(B) can be expressed as:
\begin{equation}
\rho(B) = \frac{(1+ \mu_l^2 B^2)(1+ \mu_h^2 B^2)}{\sigma_{l,0} (1+ \mu_h^2 B^2) + \sigma_{h,0} (1+ \mu_l^2 B^2)}.
\end{equation}
The magnetoresistance is then given by 
\begin{equation}
    MR(B) = \frac{\rho(B)-\rho(0)}{\rho(0)}.
    \label{eq:mr}
\end{equation}

The field-dependent magnetoresistance in Fig. \ref{fig:MR}(F400) was fit to Eq. \ref{eq:mr} for the as-grown sample. The solid curves show fits for all of the T~$\leq$~20~K data, which are in good accord with both the low-field and high-field regions. Interestingly, the two-channel model produces a linear-in-B MR at high fields, a feature observed in many other materials. 

\begin{figure}
\includegraphics[width=0.95\columnwidth]{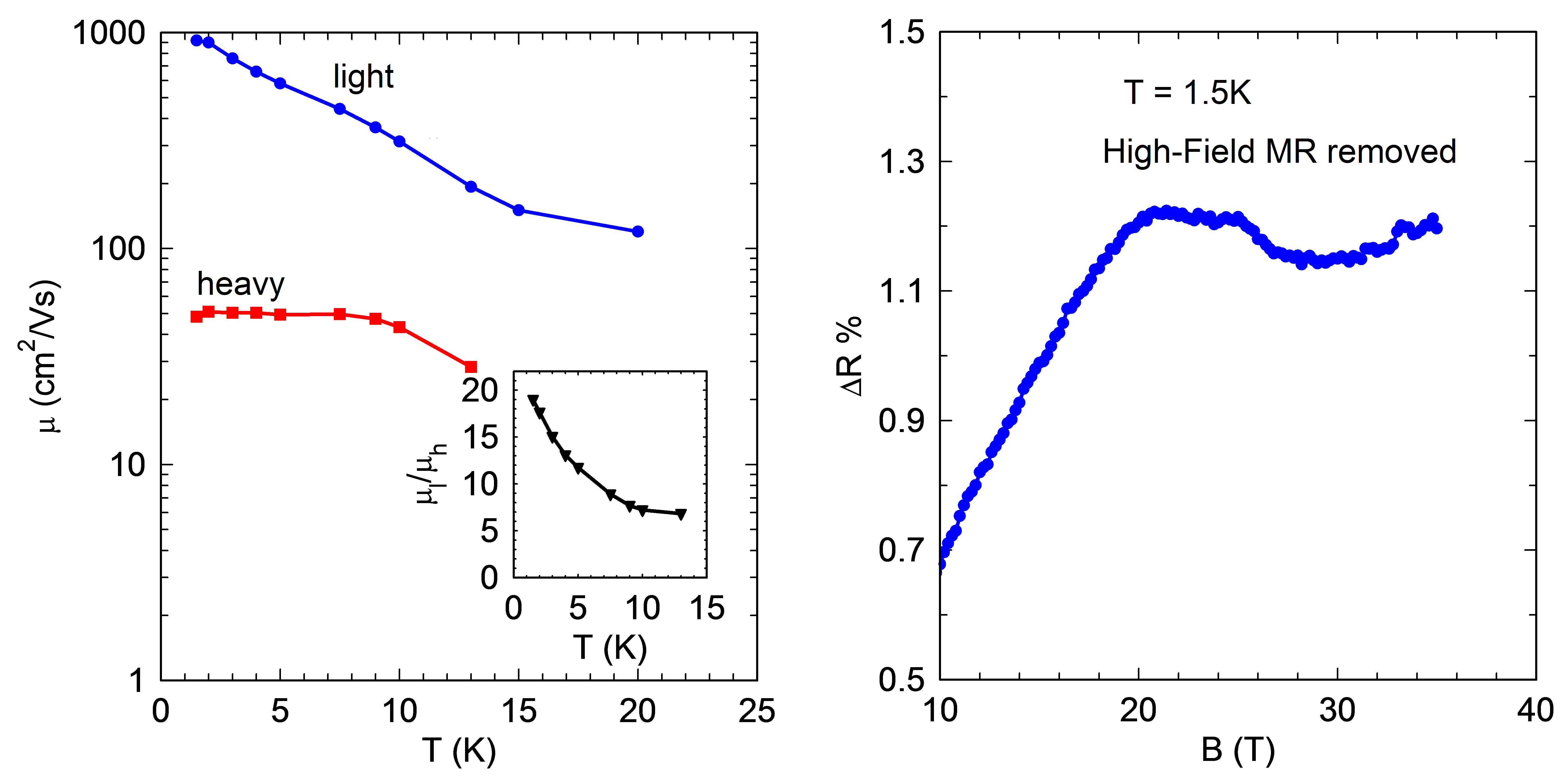}
     \caption{(left) Electron mobilities for the light (blue) and heavy (red) carriers derived from the two-carrier magnetoresistance of the as-grown CrVTiAl film. Above T~=~13~K, $\mu_l$ does not contribute significantly to the MR in the measured field region. (right) MR(B) with the high-field linear background removed, showing the onset of Shubnikov de-Haas oscillations.}
\label{fig:mu400}
\end{figure}

Figure \ref{fig:mu400} (left panel) shows the mobilities obtained from the MR curve-fits as a function of temperature for the F400 film. The mobility in both channels decreases with increasing temperature due to increased carrier scattering. At low temperatures the mobilities for the two channels differ by an order of magnitude, where the high-mobility channel has $\mu_h=920$~cm\textsuperscript{2}/Vs and the low-mobility channel has $\mu_l=48$~cm\textsuperscript{2}/Vs at T~=~1.5~K. The ratio of these mobilities is plotted as a function of temperature in the inset of Fig. \ref{fig:mu400}. These mobilities can be related to carrier effective masses within the Drude model by $\mu = e \tau / m^*$, where $\tau$ is the scattering time and $m^*$ is the effective mass.\cite{AshcroftMermin} If $\tau$ is the same for both carriers, the ratio of mobilities will be inversely equal to the ratio of effective masses. This ratio decreases from $\mu_h/\mu_l=19$ at 1.5~K to $\mu_h/\mu_l=7$ at 20~K.

\subsubsection{Additional MR features}
There are several other interesting features in the MR in addition to the two-carrier resistivity behavior, including a weak localization cusp, suppression of the quadratic MR at high temperatures, and the onset of Shubnikov–de Haas oscillations. These smaller features show both negative as well as positive MR, and occur at various temperature and field ranges.

\textit{Weak Localization Cusp} - In the F600 film a sharp MR cusp is observed for B~<~1.5~T and temperatures less than 5~K. This negative MR cusp is characteristic of weak localization (WL) arising from quantum interference of the wave-like nature of the scattering carriers. WL is well-described by the Hikami-Larkin-Nagaoka (HLN) model of quantum coherence and was used to fit the experimental cusp. The inelastic electron-electron and magnetic scattering terms are found to dominate, giving characteristic fields $B_i = 0.26$~T and $B_s = 0.39$~T. These correspond to quantum coherence lengths of $l_i = 44$~nm and $l_s = 35$~nm, respectively. \cite{Hikami1980} This strong electron-electron and magnetic scattering arises from the compensated ferrimagnetism.

\begin{figure}
\includegraphics[width=0.95\columnwidth]{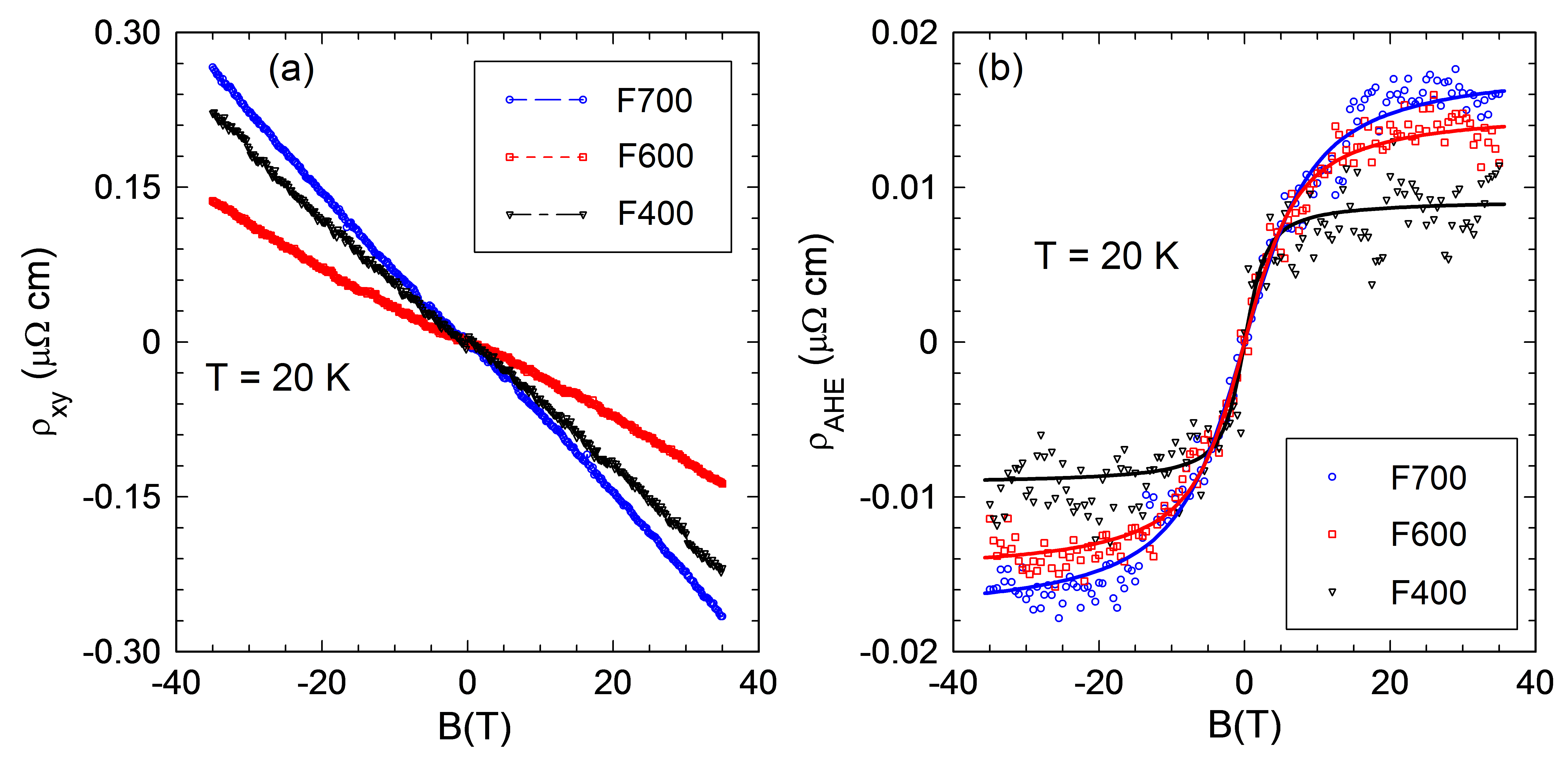}
     \caption{Hall resistivity of CrVTiAl films as a function of field measured at T~=~20~K for F400, F600, and F700. The Hall resistivity consists of (a) a linear-in-B component due to the ordinary Hall effect, plus (b) a small saturating component from the anomalous Hall effect that saturates by B~=~15~T.}
\label{fig:hall}
\end{figure}

\textit{Suppression of MR} - At higher temperatures, the dominant positive MR found in all of the films is slightly suppressed by an additional small {\it negative} MR. This suppression is present in all samples above T~=~20~K, and is strongest in the F700 sample, where the suppression begins to appear at T~=~9~K, and extends out to B~=~15~T by T~=~50~K. The suppression by the negative MR at relatively low fields is unusual in that it appears only for \textit{higher} temperatures, rather than at lower temperatures. This suppression is related to the saturating magnetization that is observed in the anomalous Hall effect (AHE). The Hall results shown later in Fig. \ref{fig:hall}(b) exhibit a small change in the Hall conductivity that saturates by B~=~15~T, a field that coincides with the disappearance of the suppression in the MR. This ferrimagnetic component in the magnetic moment would give rise to a negative MR that saturates at higher fields, which is expected and usually observed when carrier scattering decreases as the moments become aligned at high fields.

\textit{High-field Oscillations} - All three films in Fig. \ref{fig:MR} show small ripples in the MR at the highest fields and lowest temperatures. This appears to be related to the onset of Shubnikov-de Haas quantum oscillations (right panel, Fig. \ref{fig:mu400}), which are likely to be observed for the relatively high-mobility carriers with $\mu_h=920$~cm\textsuperscript{2}/Vs. The oscillations give a period of $\Delta (1/B) = 0.008~~T^{-1}$ corresponding to a Fermi surface area of $A = 1.3 \times 10^{14} cm^{-2}$. Higher fields will be required to elucidate this further.

\subsection{Hall Effect and Magnetization}
Figure \ref{fig:hall} shows the Hall resistivity $\rho_{xy}$(B). We find a linear field dependence that is associated with the ordinary Hall effect (OHE). The slope gives an effective carrier density of $n\sim10^{22}$~cm\textsuperscript{-3}, which, although high for a semiconductor, is comparable to other Heusler compounds. The effective carrier density is seen to vary with annealing by up to a factor of two. In addition to the major linear-in-B OHE, a small saturating component is present. After subtracting a linear component, Fig. \ref{fig:hall}(b) shows a clear saturating anomalous Hall effect (AHE), which is about 10~\% of the overall Hall voltage. This AHE saturation near B~=~15~T is associated with a saturating magnetic moment, consistent with observations from other compensated ferrimagnets.\cite{Nayak2015} However, the high saturation field observed in CrVTiAl is indicative of the presence of exceptionally large magnetic exchange within the material. This saturation field corresponds to the field range of the MR suppression, as noted previously. 

\begin{figure}
    \centering
    \includegraphics[width=8cm]{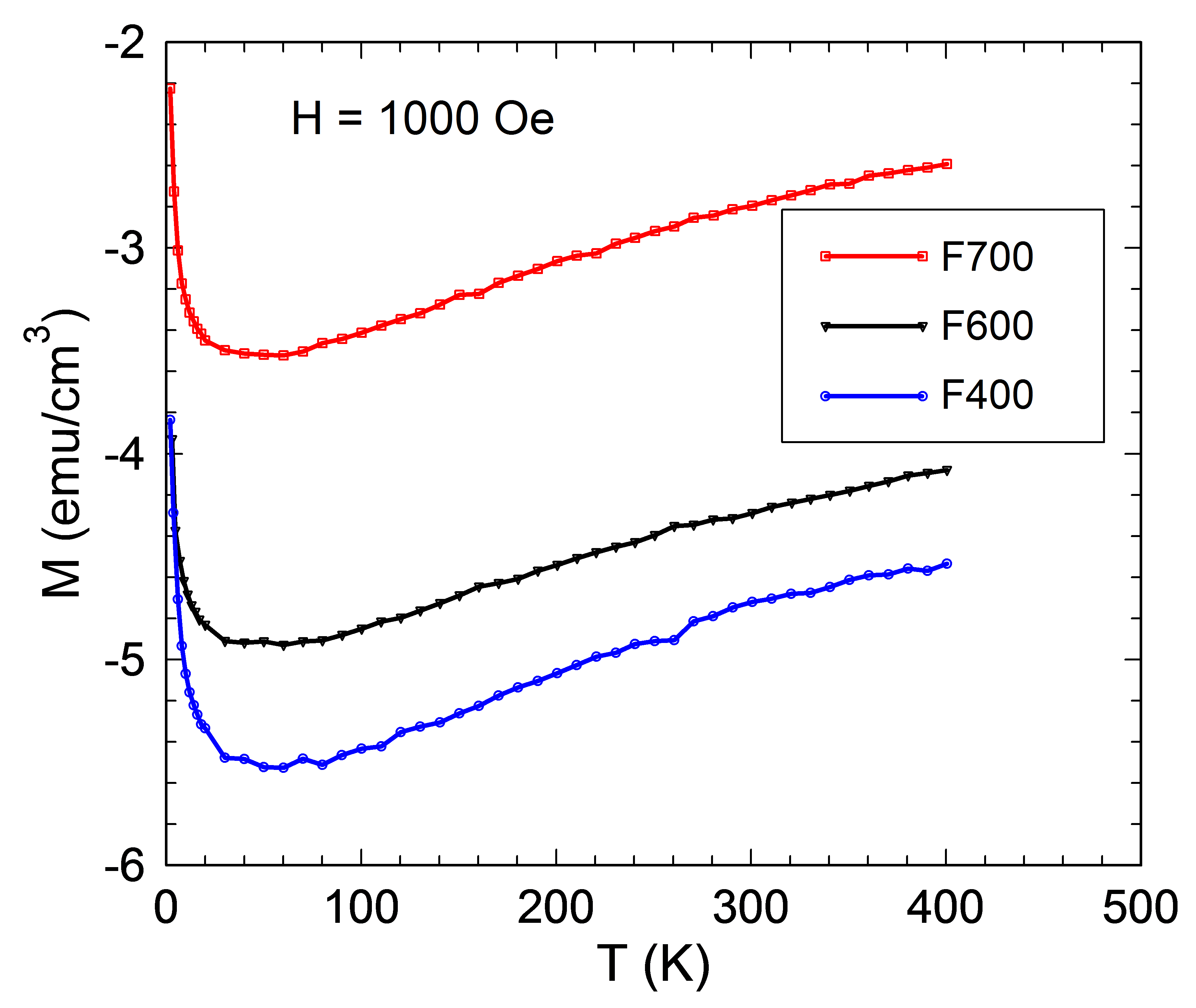}
    \caption{Relative change in magnetization of CrVTiAl as a function of temperature for the F400, F600 and F700 films, measured at a field of B~=~0.10~T. A paramagnetic component appears below T~=~50~K, but M(T) increases monotonically up to T~=~400~K. The increasing moment is indicative of a compensated ferrimagnet with T\textsubscript{C}~>~400~K. Note that the Si substrate contributes a negative temperature-independent diamagnetic offset.}
    \label{fig:mag}
\end{figure}

Figure \ref{fig:mag} shows how the magnetization M(T) changes as a function of temperature. The values are negative because of the large linear-in-B diamagnetism of the Si substrates. As diamagnetism is temperature-independent, any changes in the magnetization with temperature can be attributed to the films. M(T) is seen to increase monotonically from T~=~50 to 400~K for all three annealing conditions. There are two temperature regions, below and above T~=~50~K. At low temperatures the magnetization decreases as 1/T for increasing temperature until reaching a minimum near T~=~50~K, after which the moment increases monotonically. In compensated ferrimagnets the moment is at a minimum at the compensation point, after which it increases to a maximum before going to zero again at the Curie temperature.\cite{Stinshoff2017} This behavior stems from increased thermal fluctuations with increasing temperature. 

\section{THEORETICAL MODELING OF  ELECTRONIC AND MAGNETIC STRUCTURE}
\subsection{Ordered Atomic Arrangements}
In order to  understand our experimental results, we computed the groundstate electronic and magnetic properties of CrVTiAl alloys. In particular, we performed first-principles calculations on all the unique atomic arrangements of the quaternary Heusler structure as follows. The spacegroup of the quaternary Heusler compounds is F-43m (No. $216$), occupying Wyckoff positions $a$, $b$, $c$, and $d$. For four distinct atomic species X, X$^{\prime}$, Y, and Z arranged on the four Wyckoff sites, 4! permutations can be made. However, this number can be reduced by two constraints. Since the Wyckoff positions lie along the body diagonal of the conventional cubic cell, cyclic permutations must form an equivalence relation between permutations, resulting in six equivalence classes. Furthermore, since spacegroup $216$ belongs to cubic Bravais lattices there is no preferred cardinal direction, forcing XX$^{\prime}$YZ to be equivalent to ZYX$^{\prime}$X. We thus adduce that only three unique arrangements of atoms are possible: XX$^{\prime}$YZ, YXX$^{\prime}$Z, and X$^{\prime}$YXZ. Mapping the atomic species Cr, V, Ti and Al, to X,X$^{\prime}$,Y and Z, we obtained the groundstate electronic and magnetic structures for each of these unique atomic arrangements. We label these arrangements by their electronic phases: SFM, SGS, and metallic, which correspond to the atomic ordering along the (111) body diagonal as Cr-Ti-V-Al, Cr-V-Ti-Al, and Cr-V-Al-Ti, respectively.

Electronic structure calculations were performed using the pseudopotential projector augmented-wave (PAW) method implemented in the Vienna {\it ab initio} simulation package (VASP) \cite{Kresse1996,Kresse1993, Kresse1999} with an energy cutoff of 400 eV for the plane-wave basis set. Exchange-correlation effects were treated using the SCAN meta-GGA scheme \cite{JSun2015}, where a 18$\times$18$\times$18 $\Gamma$-centered k-point mesh was used to sample the Brillouin zone. All sites in the unit cell along with its dimensions were relaxed using a conjugate gradient algorithm to minimize energy with an atomic force tolerance of 0.05 eV/ \AA~  and a total energy tolerance of 10$^{-4}$ eV. The computed structural parameters agree with the corresponding experimental values. \cite{Medeiros2014,Medeiros2015}

\begin{figure*}
    \centering
    \includegraphics[width=2\columnwidth]{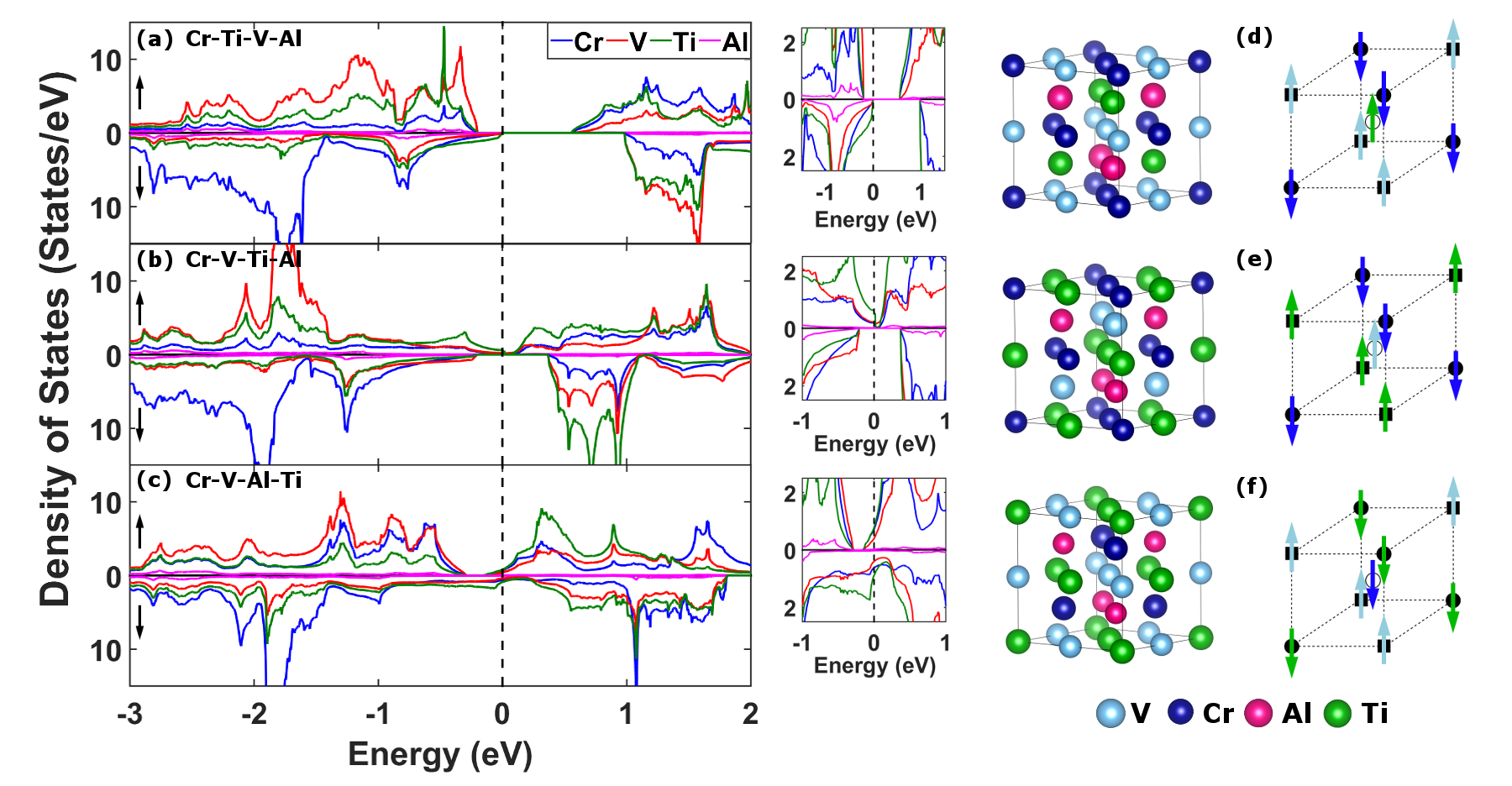}
    \caption{(a)-(c) Site-projected densities of states, and (d)-(f) crystal structures for various arrangements of Cr, V, Ti, and Al. A zoomed-in view of the region around the Fermi level is shown in the center column. Schematic representation of the cubic quasi-antiferromagnetic lattice with a compensating body-center site is shown on the far right.}
    \label{fig:dos_struct}
\end{figure*}

Figure \ref{fig:dos_struct} shows the CrVTiAl crystal and its magnetic structure along the corresponding site-projected densities-of-states (DOSs) for the three arrangements considered. Panel (a) shows the DOS for the SFM phase, exhibiting a large 0.551 eV band gap formed in the transition metal bands. Here, the valence band is dominated by V and Ti atomic character and the conduction band by Cr and Ti character. The semiconducting behavior is consistent with previous first-principles studies. The formation of the band gap in the Cr, V, and Ti hybridized band exhibits characteristics of covalent magnetism as explained by Williams {\it et al}.\cite{Williams1981}

In the SFM phase (Cr-Ti-V-Al), the opening of the band gap is facilitated by the stabilization of a fully compensated ferrimagnetic order on a tripartite lattice formed by the Cr, V, and Ti sites. \cite{Galanakis2014} The Cr and V atoms form a simple cubic cell in the quaternary Heusler structure, shown in Fig. \ref{fig:dos_struct}(d), with Ti or Al occupying the body-center position. The predicted values of the Cr, V, Ti and Al moments are -3.319, 2.665, 0.669, and 0.004 $\mu_{B}$, respectively. These values agree with previous studies. \cite{Galanakis2014, Stephen2019a} The sum of the V and Ti magnetic moments cancel the Cr spin moment, since the Al atom carries only a negligible moment. References \onlinecite{Ozdogan2013a} and \onlinecite{Skaftouros2013} discuss the zero total spin magnetic moment in the unit cell in terms of the Slater–Pauling rule.

In the phase with the arrangement Cr-V-Ti-Al along the body diagonal (Fig. \ref{fig:dos_struct}(e)), the position of Ti and V are swapped with respect to the preceding SFM case, resulting in weaker AFM exchange couplings between the neighboring Cr and Ti atoms. This arrangement yields magnetic moments on Cr, V, Ti and Al of -3.063, 1.905, 1.231, and -0.005 $\mu_{B}$, respectively. The sum of the V and Ti moments still cancels the substantial Cr spin moment, leaving a zero total spin magnetic moment in the unit cell. Due to the weakening of exchange couplings, the majority-spin channel in the DOS is metallic (Fig. \ref{fig:dos_struct}(b)).  The minority-spin channel maintains a 0.584 eV band gap, yielding a fully compensated ferrimagnetic spin-gapless semiconductor.

In the metallic phase with the atomic arrangement  Cr-V-Al-Ti along the body diagonal, antiferromagnetic correlations are further weakened by swapping Cr and Ti, see Fig. \ref{fig:dos_struct}(f). We now find a distinct change in the magnetic moments on Cr, V, Ti and Al with -1.319, 1.800, -0.435, and 0.009 $\mu_{B}$, respectively, leading to a marginal total magnetic moment of 0.3148 $\mu_{B}$ in the unit cell. The corresponding DOS in Fig. \ref{fig:dos_struct}(c) shows the disappearance of the band gap. However, a slight perturbation in the magnetic exchange splittings or atomic hybridizations could open a gap given the minimal DOS weight at the Fermi level in the minority channel. 

There is a delicate balance between the AFM and FM correlations, which drives the characteristic differences between the SFM, SGS, and metallic phases.  In the SFM phase, the two strongest magnetic elements (Cr and V) are placed on the simple cubic lattice, enabling strong AFM coupling. In the SGS and metallic phases, the magnetic atoms are substituted away to the body-center position, weakening the AFM correlations. Moreover, as the stronger magnetic atoms are placed in line with Al, FM double-exchange interactions are enhanced. The interplay of these two opposing interactions is reflected in the electronic structure through the gradual closing of the gap. These theoretical observations, along with the experimental results, show that CrVTiAl forms in the SGS phase, with Cr-V-Ti-Al arranged consecutively along the (111) direction. This structure is the quaternary equivalent of the inverse Heusler XA structure, where the chemical formula X\textsubscript{2}YZ corresponds to X-X-Y-Z along the diagonal.

A summary of the lattice, electronic, and magnetic structure along with total energies for each atomic arrangement is given in Table I.

\begin{table*}[ht!]
\caption{\label{table:ordered_summary}Comparison of various theoretically predicted  properties for the three unique atomic arrangements in pristine CrVTiAl. The three structures involve different permutations of the four sublattices along the (111) body-diagonal as discussed in the text.}
\begin{ruledtabular}
\begin{tabular}{lccc}
&Cr-Ti-V-Al & Cr-V-Ti-Al & Cr-V-Al-Ti \\
\hline
Electronic Phase            &  SFM  &  SGS  &  Metallic  \\
Lattice Parameter (\AA)     &  6.185  &  6.163  &  6.083  \\
Relative Energy (eV)        &  0.000  &  0.731  &  2.471  \\
Majority Gap (eV)           &  0.750  &    -    &    -    \\
Minority Gap (eV)           &  0.967  &  0.571  &    -    \\
Valence Band Exchange Splitting (eV)   &  0.1986  &    0.2077    &    -    \\
Electronic Gap (eV)         &  0.551  &    -    &    -    \\
Cr Moment ($\mu_{B}$)       & -3.319  &  -3.063 &  -1.319 \\
V Moment ($\mu_{B}$)        &  2.665  &  1.905  &  1.800  \\
Ti Moment ($\mu_{B}$)       &  0.669  &  1.231  &  -0.435 \\
Al Moment ($\mu_{B}$)       &  0.004  &  -0.005 &  0.009  \\
Wigner-Seitz Magnetic Moment ($\mu_{B}$)&  0.0  &  0.0  &  0.3148\\
\end{tabular}
\end{ruledtabular}

\end{table*}

\subsection{Substitutional Disorder}
Our theoretical modeling predicts that the SFM phase is energetically more favorable. However, the experimental results indicate the presences of a majority SGS phase in our thin films. To further delineate the stability of the SGS (Cr-V-Ti-Al) electronic and magnetic structure against substitutional disorder, we considered all pairwise mixings and important simultaneous substitutional disorders, see Table \ref{table:disorder} for details.
 
The effects of substitutional disorder were investigated via {\it ab initio} calculations carried out using the all electron, fully charge- and spin-self-consistent Korringa-Kohn-Rostocker coherent-potential-approximation (KKR-CPA) scheme.\cite{Bansil1979a,Bansil1979b,Bansil1981,Bansil1999,Bansil1993} The KKR-CPA approximates the average properties of the disordered system by averaging these properties  over a site embedded in a self-consistently determined effective medium. The KKR-CPA goes beyond the simpler virtual crystal approximation (VCA) and average t-matrix approximation (ATA)\cite{Gonis1992}. Moreover, KKR-CPA does not rely on {\it ad hoc} free parameters, and thus provides a more satisfactory treatment of disorder effects. Exchange-correlation effects were treated using the local spin-density approximation, (LSDA)\cite{Perdew1981} where the charge- and spin-densities were converged to $10^{-4}$ Ry. A muffin-tin radius of 2.51 a.u. was used for all 4 atomic species. The experimental value of the lattice constant of 6.136 \AA~was used.\cite{Stephen2016}

\begin{figure}
    \centering
    \includegraphics[scale=0.23]{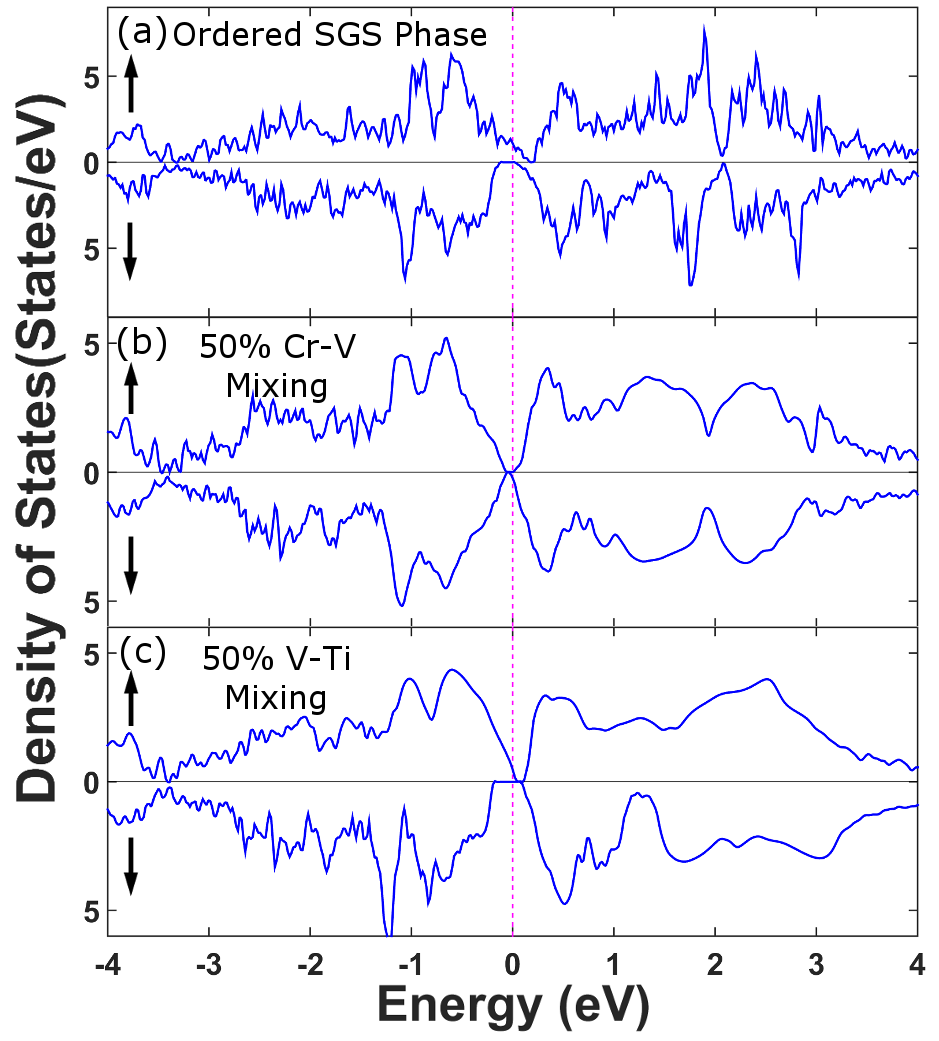}
    \caption{Total densities of states for various structures of CrVTiAl. (a) Fully-ordered structure. SGS behavior of the fully-ordered phase is preserved for the two lowest energy disordered states. (b) Cr-V mixed and (c) V-Ti mixed disordered phases.   }
    \label{fig:doskkr}
\end{figure}

Table \ref{table:disorder} compares the total energies relative to the SGS groundstate, where the magnetic moments and electronic phases (SGS, semiconducting, or metallic) of the various disordered structures are considered. Overall, most disordered states lie at energies greater than 200 meV and are predicted to be metallic, with the states above 300 meV producing a net total Wigner-Seitz magnetic moment of $0.25-0.50 ~\mu_{B}$. 

Figure \ref{fig:doskkr} compares the total DOS of the pristine SGS phase (Panel (a))\cite{footnote} to the energetically favorable Cr-V and V-Ti disordered states. Panel (b) shows the persistence of the SGS state, albeit with a reduction in the band gap. The reduction in band gap is a direct result of transforming the SFM phase to the SGS phase under Cr-V swapping, driving the system towards metallicity. Panel (c) shows a similar persistence of the SGS state but with an increased band gap. Swapping V and Ti atoms transforms the SFM phase to the metallic phase, enhancing AFM correlations. Overall, we find the SGS state to be quite robust under substitutional disorder perturbations.

In order to connect our result on the SGS phase to the MR(B) measurements in Section II.B, we estimate the relevant effective-mass ratios to provide a baseline. The effective-mass ratio in the SGS phase is calculated using $\left(m^*\right)^{-1}=\frac{1}{\hbar^2}\frac{\partial^2 E(k)}{\partial k^2}$ for k along the high-symmetry directions. We obtain a ratio of 6 - 20, within an energy window of $\pm$ 30 meV about the Fermi level to account for any intrinsic doping of the experimental sample.\cite{AshcroftMermin} The calculated mass ratios fall within the range of measured mobility ratios. A more rigorous modeling of the transport process is, however, required to give obtain deeper insight into the origin of the large observed ratio of carrier mobilities.

\begin{table*}[ht!]
\caption{\label{table:disorder}Comparison of the theoretically predicted total energies (relative to the SGS ground state), net Wigner-Seitz magnetic moments, and the electronic phases of the various disordered structures of CrVTiAl in the SGS phase. The disordered phases were obtained by exchanging 50~\% of the atoms.}
\begin{ruledtabular}
\begin{tabular}{cccc}
Atomic          & Relative      & Net Wigner-Seitz          & Electronic      \\
Substitutions   & Energy (meV)  & Magnetic Moment ($\mu_B$) & Phase         \\
 \hline
-    &    0            &   0.0   &     SGS \\
Cr-V   &    17.69  &   0.0   &     SGS\\
V-Ti   &    191.84  &   0.0   &     SGS \\
Ti-Al  &    208.17   &   0.0   &     Metal\\
Cr-V,Ti-Al& 292.52  &   0.0   &     Metal\\
V-Al   &    383.68  &   0.50   &     Metal\\
Cr-Ti  &    391.84  &   0.56   &     Metal\\
Cr-Al  &    393.20  &   -0.25  &     Metal\\
Cr-V-Ti &   424.50   &   0.46   &     Metal\\
\end{tabular}
\end{ruledtabular}

\end{table*}

Notably, our first-principles calculations predict that the SFM phase in which the magnetic atoms are arranged in the order Cr-Ti-V-Al along the (111) body diagonal, is energetically the most favorable phase. However, the experimental results indicate the presence of a majority SGS phase in the thin films. Our first-principles results indicate that the atoms arranged along the body diagonal in the SGS phase will lie in the order Cr-V-Ti-Al. However, to obtain the fully compensated SFM phase, the magnetic atoms must arrange in the order Cr-Ti-V-Al along the body diagonal; the weaker magnetic atom (Ti) would then lie in the center of the $\it{bcc}$ cube that is created by the two strongly-magnetic atoms (Cr and V). With increasing disorder, as the larger moments are placed in the $\it{bcc}$ center, the system transforms into the SGS phase, and with further disorder it transforms into the metallic phase. 

\section{SUMMARY}

In the investigated CrVTiAl films, we have shown the presence of two parallel conducting channels in both $\rho$(T) and MR(B) measurements.  The results of $\rho$(T) measurements indicate a two-channel semiconducting SGS character. The gapless channel is found to have a small overlap of $\sim$~0.05~eV, while the gapped channel has an activation energy of $\Delta E$=~0.1~-~0.2~eV. The results of M(B) measurements identify high- and low-mobility carriers having a large mobility ratio varying from $\mu_h/\mu_l$~=~7~to~19 in reasonable accord with our theoretical estimate of the effective mass ratio in the SGS phase. We have investigated the electronic and magnetic structures for a variety of ordered and disordered phases of CrVTiAl alloys using self-consistent first-principles calculations. Our results suggest that in order to successfully design a room temperature spin filter, it will be important to investigate other quaternary compounds to determine which set of atoms will more easily form the desired SFM phase.

\section{Acknowledgements}
The experimental work was supported by the National Science Foundation grant ECCS-0142738. A portion of this work was performed at the National High Magnetic Field Laboratory, which is supported by the National Science Foundation Cooperative Agreement No. DMR-1644779 and the State of Florida.  The theoretical work was supported by the US Department of Energy (DOE), Office of Science, Basic Energy Sciences grant number DE-FG02-07ER46352 (core research), and benefited from Northeastern University's Advanced Scientific Computation Center (ASCC), the NERSC supercomputing center through DOE grant number DE-AC02-05CH11231, and support (testing the efficacy of advanced functionals in complex materials) from the DOE EFRC: Center for Complex Materials from First Principles  (CCM), under DOE grant number DE-SC0012575.

\onecolumngrid

\clearpage

\clearpage

\bibliographystyle{apsrev4-1}
\bibliography{References}

\begin{thebibliography}{40}
\bibitem [{foo()}]{footnote}%
  \BibitemOpen
  \href@noop {} {}\bibinfo {note} {The calculations in Fig.
  \ref{fig:dos_struct}(b) were performed using the new SCAN meta-GGA functional
  which captures stronger AFM coupling, as compared to those predicted by LDA
  in Fig. \ref{fig:doskkr}(a), along with various local onsite potential
  renormalizations. Together, this results in the band shifts observed when
  comparing Figs. \ref{fig:dos_struct}(b) and \ref{fig:doskkr}(a)}\BibitemShut
  {NoStop}%
\end{thebibliography}

\end{document}